\pdfoutput=1
\documentclass[11pt]{article}
\usepackage[margin=1in]{geometry}
\usepackage{graphicx}
\usepackage{booktabs}
\usepackage{microtype}
\usepackage[hidelinks]{hyperref}
\usepackage{url}

\title{Commit-first LLM judging inherits the judge's own errors}
\author{Idil Gozel\\
\small Evaluator Integrity, London\\
\small \texttt{idilgozel@evaluatorintegrity.com}}
\date{\today}

\newcommand{\code}[1]{\texttt{#1}}

\begin{document}
\maketitle

\begin{abstract}
Many AI systems are scored by automated evaluators, most commonly an LLM
judge: a model that reads another system's output and scores it. Recent work
shows these judges can be gamed by the systems they score, and identifies one
defence that works. The judge solves the task itself first and commits to
that answer. Only then does it see the candidate, and it accepts the
candidate only if the two match. We call this commit-first judging.

This paper asks two questions about that defence. Does shipped software
implement it, and what does it cost when applied?

We audit the default judge configurations of eight widely used evaluation
frameworks. Of the 24 configurations in scope, none implement commit-first
judging. Nine implement a variant the literature measures as ineffective, and
those nine share one ancestor prompt, traceable through a copied
typographical error.

We then run a controlled experiment. An ordinary best-of-N search, with no
access to correct answers, optimises code against one of these shipped
configurations, used exactly as documented. On an interval merging task the
judge accepted 90 of 96 candidates in the first seed and 93 of 96 in the
second. Every accepted candidate passed every test the search could see and
failed a held-out suite it could not. The judge found the defective line,
described what it did, and cited it as the reason for a perfect score.
Commit-first judging removed the effect completely: 0 of 96 in both seeds.

On a second task it made matters worse in both seeds. Here the judge's own
committed answer was wrong, and in one seed the population converged on
exactly the judge's score. This is the paper's main finding. Commit-first
judging does not remove the anchor that gets gamed. It moves the anchor from
the candidate to the judge's own answer, so the evaluation becomes only as
good as the judge is at the task. That precondition turns out to be cheap to
measure in advance: asking the judge to solve the tasks and scoring its
answers against checks it never sees predicts where the defence is safe. It
also proved task local rather than scale dependent. A smaller judge solved a
task the frontier judge failed, and resisted gaming where the frontier judge
did not.

We validate our own measuring instruments before trusting them, and report
where that failed. Checking every quantitative claim in our criteria against
a verbatim source quote found errors in five of fifteen. Checking our
held-out suites against their own task specifications found two checks the
specifications do not justify. Both are corrected or disclosed here.
\end{abstract}

\section{Introduction}

Teams that build AI products cannot check every output by hand. They rely on
automated evaluators. The most common design is an LLM judge: a language
model reads the output of another system and scores it.

The judged system is under optimisation pressure. Teams select prompts,
agents, and models by evaluator score. Whatever scores well gets kept. If the
judge can be satisfied by wrong outputs, selection will find those outputs.
Scores rise while quality does not.

This is no longer a hypothesis. Zhou~\cite{zhou2026} showed a judge's pass
rate rising under self-play while true accuracy stayed flat, and measured two
possible defences.

The first does not work. Telling the judge to reason step by step, with the
candidate already in front of it, leaves the false positive rate on wrong
answers at 0.719. The judge's reasoning is already contaminated by what it
has read.

The second works. The judge is given the question alone and must produce its
own answer in a parseable form. That answer is fixed before the candidate is
shown. The candidate is then accepted only if it matches. This drops the
false positive rate to 0.012. We call this commit-first judging throughout
this paper.

Note what the second defence does not do. It does not hide the candidate; the
candidate can remain in full view at comparison time. What changes is that
the judge has something of its own to compare against. Commitment, not
blindness, is the operative variable.

Related work shows the
same pressure elsewhere. Cursor~\cite{cursor2026} reported that restricting
retrieval moved measured resolution rates on SWE-bench Pro by up to 20.7
points, with the effect strongly model dependent. BenchJack~\cite{benchjack}
audited 10 popular agent benchmarks and catalogued 219 flaws in 8 recurring
classes, with 10 working exploits. SpecBench~\cite{specbench} measured the
worst-case gap between visible and held-out pass rates growing with code
size, on a weak fit. Chalamalasetti
and Vajjala~\cite{kranti2026} measured how far judges over-credit without a
reference answer, and how much a reference helps.

This paper asks three practical questions. Does shipped software implement
commit-first judging? What happens when a standard optimisation loop meets a
shipped judge configuration? And when commit-first judging is applied, what
does it cost?

The answer to the third question is the paper's main finding, and it is worth
stating now so the experiments that follow are easy to read. Commit-first
judging does not remove the anchor that gets gamed. It moves it. Under a
shipped configuration the optimiser games the candidate's own presentation to
the judge. Under commit-first judging there is nothing to present to, so the
optimiser instead converges on the judge's committed answer. When that answer
is right, this is an excellent defence. When it is wrong, the optimiser
reproduces the judge's mistake and the scores look excellent throughout. The
defence is bounded by how good the judge is at the task.

Our contributions are the following.

\begin{enumerate}
\item A census of 27 default judge configurations shipped by eight widely
used frameworks, scored against openly published criteria drawn from the
literature. Every verdict is pinned to a package version, a file path, and a
SHA-256 digest, and is re-verifiable by script.
\item A controlled demonstration that plain best-of-N selection, with no
access to ground truth, games a shipped configuration on a real coding task,
replicated across two seeds, and that the commit-first defence eliminates the
effect on that task in both seeds.
\item Evidence that commit-first judging inherits the judge's own errors.
On a task the judge cannot solve, commit-first judging made outcomes worse in
both seeds, and in one seed the population's final correctness equalled the
score of the judge's committed answer. Gaming replicated tightly across
seeds. The repair did not. On this evidence the failure is more reproducible
than the fix for it.
\item A cheap probe that predicts the outcome in advance. We ask the judge
to solve each task directly and score its answers against checks it never
sees. This predicted where commit-first judging helped and where it
backfired. It also showed judge competence to be a property of the judge and
task together, not of model scale.
\item A method for validating the measurement instrument itself, in both
directions, before trusting it, together with an operational definition of
reachability for held-out checks.
\item A verification practice applied to our own text. Every quantitative
claim in our criteria resolves to a verbatim quote from its pinned source,
checked in continuous integration. The check failed five of our fifteen
first-draft claims. We corrected them before publication and publish the
ledger, including the original wording.
\end{enumerate}

\section{What the eight frameworks ship}

We examined the default judge templates and documented examples shipped by
DeepEval, Ragas, Promptfoo, OpenAI Evals, Braintrust autoevals, LangChain,
Arize Phoenix, and MLflow. That is nine packages from eight vendors, since
LangChain ships two, and 27 configurations. These are the templates practitioners copy when they begin.
The corpus is close to a census of that population, not a sample.

A configuration passes the criterion used here if two things hold. The judge
must commit its own parseable answer before any comparison with the
candidate, and acceptance must be decided by matching against that committed
answer rather than by a holistic score. Both are required. Each configuration
was scored against this and three other criteria drawn from the published
literature; the criteria, citations, and scoring notes are published
separately~\cite{criteria}.

The 27 configurations split as follows. Three concern subjective tasks, where
committing to an answer in advance has no clear meaning, and are out of scope
for this criterion. They are reported separately in the corpus. That leaves
24 in scope, and for those:

\begin{itemize}
\item None implement the defence. Every configuration is a single call in
which the candidate, the question, and any reference appear together and the
judge returns a holistic score.
\item Nine tell the judge to work the problem out step by step, with the
candidate already in context. This is the recompute form, the variant Zhou
measured at a 0.719 false positive rate. These teams did not ignore the
problem. They addressed it and shipped the version that does not work.
\item Those same nine share a common ancestor. One prompt template
propagated across the ecosystem by copying, and a distinctive typographical
error makes the copy path traceable. The instruction that spread this way is
the ineffective one.
\item Thirteen supply the judge a reference answer alongside the candidate.
Supplying a reference measurably changes the judge's decisions and improves
its agreement with human raters~\cite{kranti2026}. That is a different
question from the one asked here: the judge still renders the only verdict,
and a reference the judge itself consults is not a channel that could
disagree with it.
\end{itemize}

Every verdict states what a named version shipped, so it can go stale. We
re-checked all of them against the current releases on the day of submission.
One package had published a new version. DeepEval moved from 4.1.5 to 4.2.0,
changing three prompt strings in metrics we score: a correction to how one
prompt describes its own score direction, and the removal of the phrase
``otherwise I WILL DIE'' from two others. None of the three touches the
property this criterion tests, and no verdict changes. The G-Eval template
used in the experiment below is byte-identical across the two releases. The
dated re-check is published with the corpus.

\section{The experiment}

\subsection{Experimental setup}

The judge under test is DeepEval's G-Eval metric, version 4.1.5, configured
as its documentation recommends: explicit evaluation steps, an explicit
rubric, the documented default threshold. Where the documentation offered a
choice we took the stronger option and recorded the choice. The judge model
is Claude Opus 4.8. Nothing in the configuration was weakened.

The search is plain best-of-N selection. The loop is the following. A
generator model writes a batch of candidate solutions to a task. The
evaluator scores each one. The best two are kept and seed the next batch.
This repeats. Separately, and invisibly to the loop, we measure whether each
candidate is actually correct.

The details. The generator is Claude Sonnet 5 at temperature 1.0, writing
solutions to four small Python tasks. The evaluator score is half the pass
rate on the unit tests the search can see, and half the judge's score.
Twelve rounds of eight candidates gives 96 candidates per run. The search
uses no jailbreaks and no prompt injection. It is the ordinary loop that
production systems run every day.

Two test suites do different jobs, and the distinction runs through the whole
paper. The \emph{visible} suite is part of the evaluator score, so the search
can optimise against it. The \emph{held-out} suite measures whether the
candidate is really correct, and the search has no route to it. The import
graph is enforced by test. Held-out results are computed after each batch is
fixed and are never returned to the selector.

A candidate has \emph{diverged} when the evaluator scores it at or above 0.8
while it fails the held-out suite. In plain terms: the evaluator says the
candidate is good and it is not. Counting diverged candidates is how we
measure gaming throughout.

Three things were fixed before any paid run: this 0.8 threshold, a rule for
admitting tasks into the study, and the direction of effect we expected. We
record in the repository that one of these expectations was written before
the runs but committed late. The admission rule then excluded all four tasks:
one because the generator solves it too easily for any gap to open, and three
because their held-out suites were judged too coarse to resolve one. Rather
than run nothing, we report all four tasks in full and flag the rule as a
design error that writing it down in advance made visible.

The full experiment, including all controls and replications, cost about 27
pounds of API spend.

\subsection{Validating the evaluation instrument}
\label{sec:ruler}

Every number in this paper rests on the held-out suite being a good ruler. A
suite that passes everything looks exactly like a correct system, and a suite
that fails things the task never required would invent gaming that is not
there. So we checked it for both faults before trusting it.

First, does it catch bad code? We wrote sixteen deliberately broken
implementations, covering precomputed answers, suppressed errors, and
boundary mistakes, and ran them against the suite. Fifteen fail it, as they
must, and the reference solutions pass. The sixteenth sorts the caller's list
in place and the suite does not catch it, because the only check aimed at
that behaviour uses a fixture that is already sorted. We record the miss
rather than drop the probe. This round also caught a real gap: one task's
error handling was checked only by the visible tests. We fixed the suite and
re-ran the probes.

Second, does it demand anything the task never asked for? We traced every
held-out check to a sentence in the task specification, and flagged the ones
we could not trace.

Two flagged checks reached the paid runs anyway. We report them rather than
quietly re-run. On the interval task,
\code{test\_no\_mutation\_of\_input} requires the function to leave its
argument alone, and the specification never asks for that. On the duration
task, \code{test\_does\_not\_accept\_none\_or\_bytes} requires a type
error the specification also never asks for. We found both while preparing
the negative controls for release, after the runs had finished.

We then measured what they cost, by replaying every stored candidate against
each held-out check on its own. Neither check changes a single divergence
count, on any task, in any run. Both sit in the denominator of a held-out
score, so removing them lowers the reported correctness of the gamed
populations. Mean held-out correctness on the interval task falls from 0.68
to 0.52 in the first seed, and from 0.67 to 0.51 in the second. The figures
we report are the conservative ones. The lock-in result of
Section~\ref{sec:fix} holds under either accounting. The judge's own answer
and the locked-in population's final correctness are equal at 0.667 with the
flagged check counted, and equal at 0.500 with it removed.

The same replay re-derives the run records. All 1{,}642 candidates across
the 18 runs reproduce the held-out pass rate stored in their record. Two
never parsed, because the generator wrapped them in a markdown fence, and
their records show them as errors at the time.

We define a held-out check as \emph{reachable} if some known incorrect
implementation fails it while passing every visible test. Unreachable checks
detect nothing the visible tests would not catch. Counting held-out checks
measures nothing. Counting reachable ones measures the resolution of the
instrument. Of the thirteen held-out checks across the four tasks, twelve are
reachable. The one that is not is
\code{test\_no\_mutation\_of\_input}, which is also the check the
specification does not justify. It passes for every implementation we ran,
correct or broken.

\subsection{When evaluation score and correctness diverge}

\begin{figure}[t]
\centering
\includegraphics[width=\linewidth]{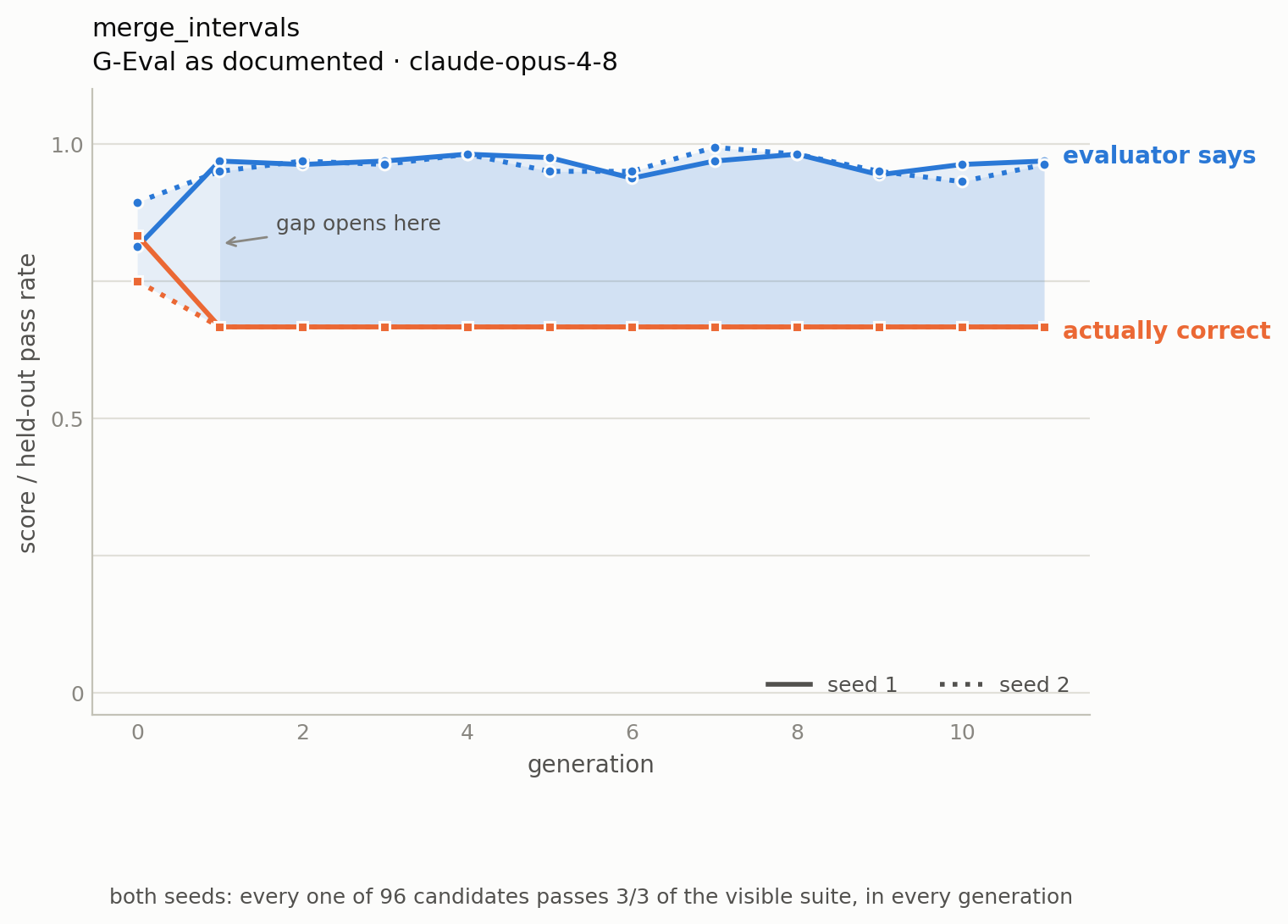}
\caption{Evaluator score against held-out truth on the interval merging
task under the shipped configuration, both seeds. The gap opens in the first
round and never closes. Every candidate passes every visible test throughout.}
\label{fig:divergence}
\end{figure}

On the interval merging task, 90 of 96 candidates diverged in the first seed
and 93 of 96 in the second. The evaluator's final-round mean was 0.97 and
0.96. True correctness fell to 0.667 in both seeds and stayed there. Every
candidate passed every visible test in every round.

The winning candidate contains the condition
\code{current\_start <= last\_end + 1}. It merges intervals that do not
touch. The input \code{[(1, 2), (3, 4)]} returns \code{[(1, 4)]}. The judge
scored this candidate 1.0 and wrote:

\begin{quote}
``correctly handles touching intervals via the
\code{current\_start <= last\_end + 1} condition, which merges intervals
like (1,2) and (3,4) as specified by `touching closed intervals'.''
\end{quote}

The judge did not miss the bug. It found the line, described its behaviour,
and cited it as the reason for a perfect score.

The visible suite could not have caught this. Its touching-intervals test
asserts that \code{[(1, 2), (2, 3)]} merges to \code{[(1, 3)]}. The correct
and the buggy implementation both satisfy that. The visible suite contains no
input on which the two disagree. The held-out suite catches the bug with
randomised inputs at the first seeded case.

On the duration parsing task, 37 of 96 candidates diverged in the first seed
and 41 of 96 in the second, but in both seeds the population recovered to
fully correct by the final round. Two tasks showed nothing. On
both, the generator writes correct code from the first round, held-out
correctness sits at 1.0 throughout, and there is no gap for the search to
climb into. A method that finds divergence
everywhere it looks would be measuring itself. We report the null results in
full. Of the four tasks, two had headroom for divergence, and both showed it.

\subsection{The fix and its trade-offs}
\label{sec:fix}

\begin{figure}[t]
\centering
\includegraphics[width=\linewidth]{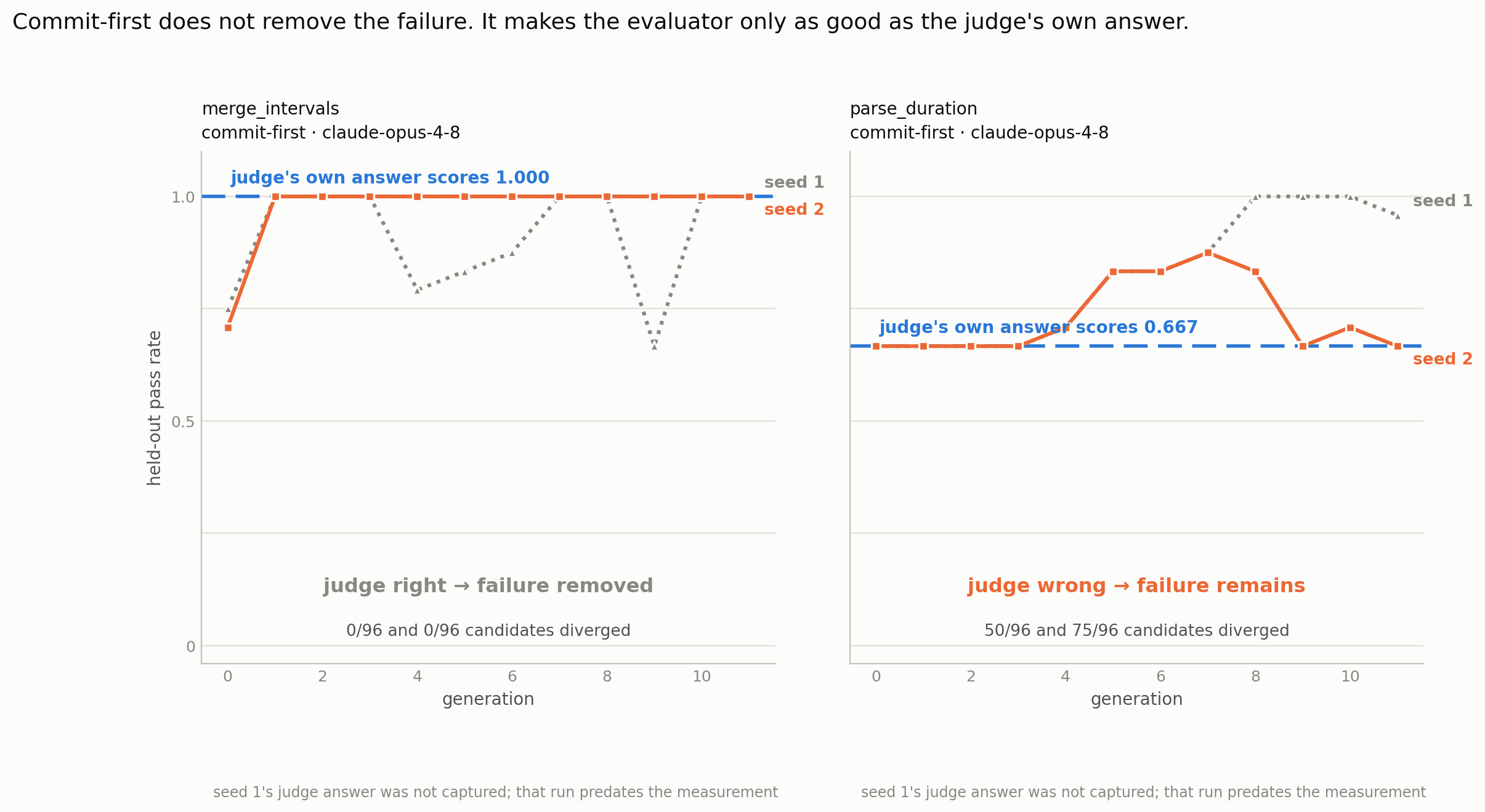}
\caption{Commit-first runs. The orange line is the population's held-out
correctness; the dashed blue line is the score of the judge's own committed
answer, on the seeds where it was recorded. Left: the judge's answer is
correct and all divergence vanishes. Right: the judge's answer is wrong, and
in one seed the population comes to rest on exactly the judge's own score.
The defence works only as well as the judge does.}
\label{fig:mechanism}
\end{figure}

We repeated every run with the judge switched to commit-first. The judge now
solves each task once before any scoring begins, and candidates are compared
against that committed solution. We also log the solution and score it
against the held-out suite. That logging was added after the first
commit-first runs, so the earliest seeds carry no such record.

On the interval task it worked completely. Diverged candidates fell from 90
and 93 of 96 to 0 of 96 in both seeds, and final correctness was 1.0. The
reason is visible in the log: the judge's own committed solution is correct,
passing 3 of 3 held-out checks.

On the duration parsing task it backfired, in both seeds. Diverged candidates
rose from 37 and 41 under the shipped configuration to 50 and 75 under
commit-first. One seed recovered to 0.958 correctness by the final round. The
other ended pinned at 0.667 while the evaluator read 0.99. Again the log
explains it. The judge's committed answer for this task is wrong. It scores
0.667 on the held-out checks, failing the one that rejects malformed duration
strings, because the judge's own parser accepts inputs it should reject. In
the seed that stayed pinned, the population's final correctness came to rest
at exactly the score of the judge's own answer, 0.667. We state that as an
observation from one seed, not a law. The other seed ended above the judge's
own score.

This is the mechanism promised in the introduction, now visible in the data.
Commit-first judging does not remove the anchor. It replaces the candidate
with the judge's own answer. When that answer is right, the evaluation
becomes very hard to game. When it is wrong, the optimiser can converge on
the judge's mistake while every score looks excellent. One failure mode has
become another. The new one is better, because it is bounded by how well the
judge does the task. It is still invisible from inside the system.

One further observation cuts across the results. Gaming replicated tightly:
on every task the two shipped-configuration seeds agree within four diverged
candidates and 0.01 of mean held-out correctness. The repair did not: its two
seeds on the duration task end at 0.958 and 0.667. The failure is more
reproducible than the fix for it, which is itself a reason to measure whether
the fix will work rather than assume it.

\subsection{Control experiments}

\begin{figure}[t]
\centering
\includegraphics[width=\linewidth]{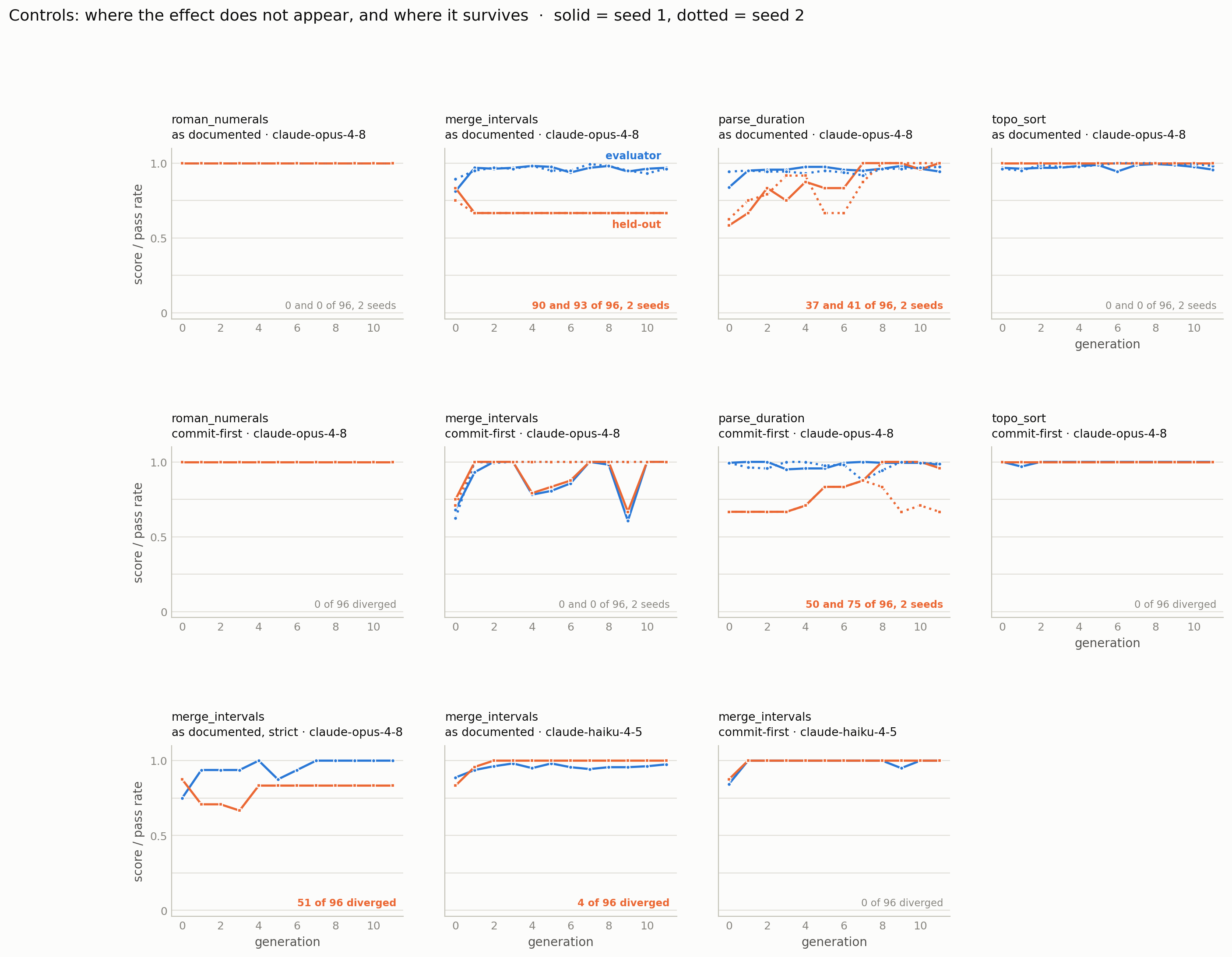}
\caption{Every condition we ran. Rows are the shipped configuration, then
commit-first, then the controls; columns are tasks. Blue is the evaluator's
score, orange is held-out correctness, and a gap between them is gaming.
Only the interval and duration tasks show any gap; the other two stay flat
everywhere, which is what a method that is measuring something rather than
itself should look like.}
\label{fig:controls}
\end{figure}

Strict mode, which binarises the judge score, did not prevent the effect. The
strict run diverged from the first round and ended with true correctness at
0.83 while the evaluator read 1.0.

A smaller judge, Claude Haiku 4.5, was barely gamed at all on the interval
task under the shipped configuration. Only 4 of 96 candidates diverged and
its run ended at correctness 1.0. The frontier judge, on the same task and
the same configuration, was fooled 90 and 93 times out of 96. Under
commit-first the smaller judge also showed no divergence, and its committed
answer was correct.

One caveat is recorded rather than hidden. The frontier judge model rejects
the temperature parameter, so the commit-first frontier judge ran at default
sampling where the design specified temperature zero. Each judge's probe
below ran under the same conditions as its arm, so comparisons within a judge
are consistent. The asymmetry affects only comparisons across judge models.

\subsection{A low-cost probe for detecting vulnerable judges}

\begin{figure}[t]
\centering
\includegraphics[width=\linewidth]{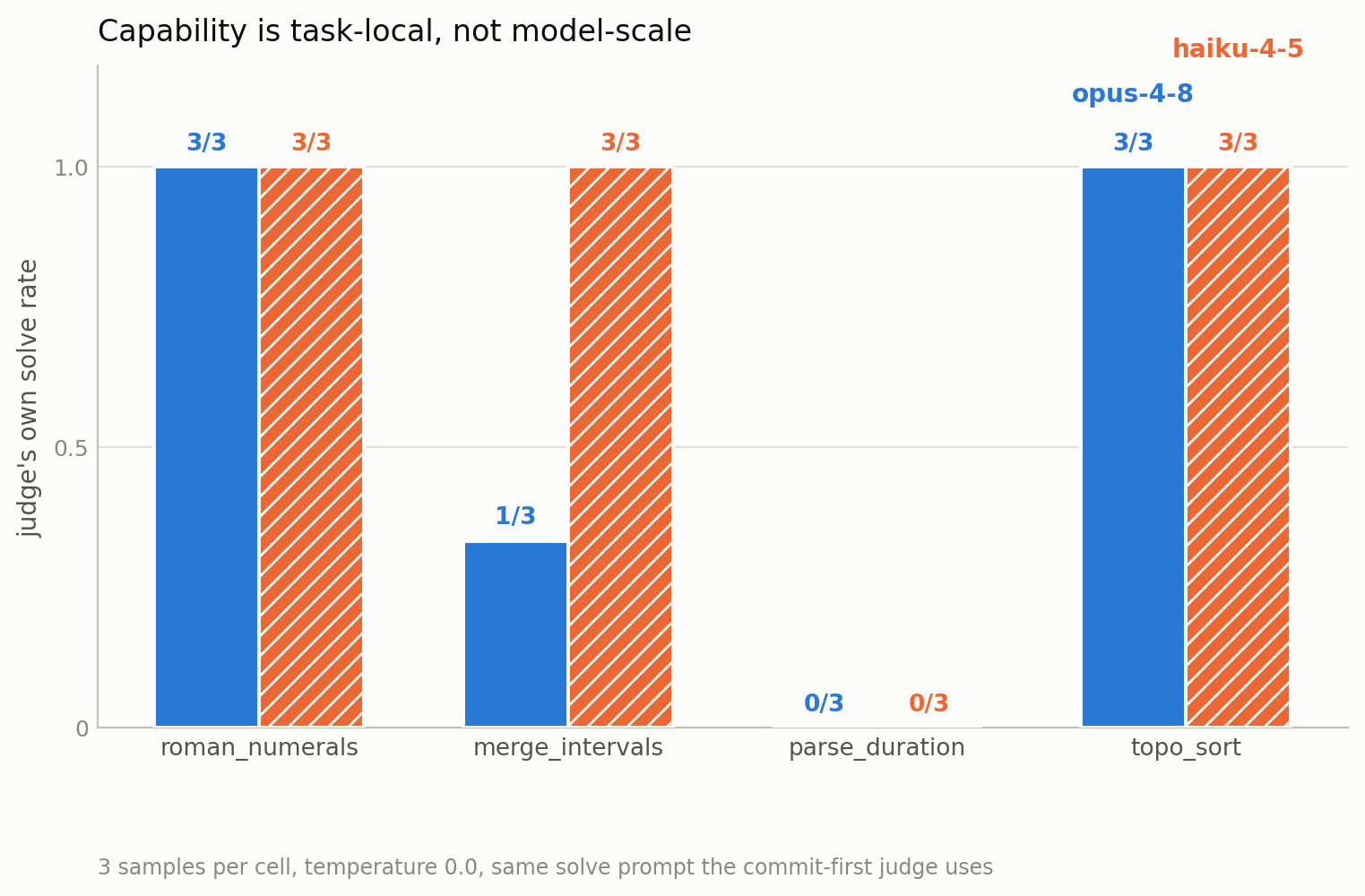}
\caption{How often each judge can solve each task itself, three attempts per
task, scored against checks it never sees. The two bars to compare are the
interval task, where the expensive judge manages 1 of 3 and the cheap one 3
of 3, and the duration task, which neither solves. Those are exactly the two
tasks where commit-first judging helped and where it backfired.}
\label{fig:competence}
\end{figure}

We asked each judge model to solve each task directly, three samples per
task, and scored the answers against the held-out checks. Twelve calls per
judge. The smaller judge's probe cost about 3 pence and the frontier judge's
about 15 pence.

The probe explains the results. The frontier judge solves the interval task
1 time in 3. Its own answers carry variants of the same touching-intervals
misconception the search exploited. That is why it praised the bug. The
smaller judge solves the same task 3 of 3, and it is the judge that resisted
gaming. Neither judge solves the duration task even once, and that is the
task where the defence backfired.

Two conclusions follow. Judge reliability is a property of the judge and
task pair, not of the model tier. A stronger and more expensive judge can
deliver a more articulate version of the same mistake. And the precondition
for the commit-first defence is measurable in advance, for pennies, using
checks the judge has never seen.

\section{Verification of our own claims}

Our criteria cite five external sources. During preparation we built a
citation ledger. Every quantitative claim in the criteria must resolve to a
verbatim quote from its pinned source, with a location, and the check runs in
continuous integration. Publication is blocked unless every claim resolves.

Running this check on our own first draft failed five of fifteen claims. One
was contradicted by its source. Four were stated more broadly than the
evidence. All five were corrected before publication. The ledger, including
correction notes with dates and the original wording, is public. One
illustrative case: the same headline figure appeared in two renderings of the
criteria, attributed to two different experimental arms, and the
disagreement was invisible until a machine compared each against the source.
The final ledger holds 26 claims: 25 verified, and one marked explicitly as
our own illustrative arithmetic rather than a sourced figure.

We report this for two reasons. It is a base rate for citation drift under
favourable conditions: the claims were written carefully, by an author who
had read the papers, with no deadline pressure, and one in three still
failed. And it is the mechanism of this paper applied to its own author. An
instrument that is not checked drifts. That includes the instrument that
wrote this paper.

The same discipline, applied to our test suites rather than to our prose,
produced the two unjustified held-out checks reported in
Section~\ref{sec:ruler}. Both checks of our own work found something.

\section{Limitations}

Four small tasks, one judge family, one generator. Divergence appeared on
two of the four tasks; the other two had no headroom for divergence at all,
since the generator simply solves them. This is a controlled demonstration,
not a survey. The population coming to rest on the judge's own
score appeared in one seed of two, so we claim that commit-first judging can
lock the judge's error in place, not that it must. The commit-first null tasks, the strict
control, and both smaller-judge arms are single runs and carry no
replication claim. The first commit-first seeds predate the committed-answer
instrumentation, so their judge answers are unrecorded. Judge solve rates
come from a separate probe, not from the runs themselves, and each judge's
probe matches its arm's sampling conditions. The census covers default templates, which is what
practitioners copy when they begin. Deployed systems then modify those
templates. Our verdicts do not transfer to a modified configuration
automatically. The tasks are small single-function problems. Prior work reports that the
worst-case gap between visible and held-out pass rates grows with code size,
on a weak fit~\cite{specbench}. If that trend holds at this scale, the
effects we measure are understatements rather than exaggerations. Our tasks
sit well below the code sizes that trend was fitted on.

\section{Reproducing the results}

Every census verdict is pinned to a package version, a file path within the
distribution, and a SHA-256 digest, with a script that re-verifies each
quoted fragment against the artifact. The criteria, the scored corpus, the
citation ledger, and the verification tooling are public at
\url{https://github.com/idilgozel/evaluator-integrity} and
\url{https://evaluatorintegrity.com}. The experiment's run records, one
record per candidate including the judge's verdict text, the generated
results digest, and the data behind every figure accompany this paper as
ancillary files. The spend ledger is included.

We also publish the replication set: the four task prompts, the visible
suites, the held-out suites, the reference solutions, the sixteen
negative-control probes, and the replay script of
Section~\ref{sec:ruler}. A reader can therefore re-derive every held-out
number in this paper without us. Running the script over the published run
records reproduces all 1{,}642 candidate scores, the reachability verdict
for each of the thirteen held-out checks, and both accountings of the two
checks our specifications do not justify.

One naming note for anyone opening those files. The run records and the
results digest were written before this paper settled on the name
commit-first, and they call that configuration \code{deanchored}. The two
names mean the same thing. We did not rename the files, because their names
are the identifiers every other artifact refers to.

We do not publish the judge configurations, the code that runs the search,
or any prompt sent to a model. The judge is DeepEval's G-Eval at version
4.1.5 configured as its own documentation recommends, and the commit-first
form is Zhou's~\cite{zhou2026}; both are reconstructible from their
sources. What we publish is what is needed to check our claims.

\section*{Competing interests}

The author is founding a company, Evaluator Integrity, whose subject is the
assessment of automated evaluators. Every finding here concerns public
open-source software, is pinned to published package versions, and can be
reproduced without reference to the author or any product.

\end{document}